\documentclass[11pt]{article}

\usepackage[bitstream-charter,greekuppercase=upright,cal=cmcal]{mathdesign}
\usepackage[T1]{fontenc}

\usepackage{amsmath,amsthm,mathabx}
\usepackage[normalem]{ulem}
\usepackage{multirow}
\usepackage{units}
\usepackage{graphicx}

\usepackage[affil-it]{authblk}
\usepackage[english]{babel}

\usepackage[margin=2.2cm]{geometry}
\usepackage{url}
\usepackage{verbatim}
\usepackage{xcolor}
\definecolor{dullmagenta}{rgb}{0.5,0,0.4}
\definecolor{dullblue}{rgb}{0.0,0,0.86}
\usepackage{hyperref}
\hypersetup{colorlinks=true,citecolor=dullblue,linkcolor=dullblue,filecolor=dullblue,urlcolor=dullblue,breaklinks=true,bookmarksnumbered=true, bookmarksopen=true,bookmarksopenlevel=1}
\allowdisplaybreaks

\newtheorem{definition}{Definition}[section]

\usepackage[noindentafter]{titlesec}

\titleformat*{\section}{\large \bfseries}
\titleformat*{\subsection}{\normalsize\bfseries}

\titlespacing\section{0pt}{12pt plus 4pt minus 2pt}{2pt plus 2pt minus 2pt}

\usepackage[backend=biber,sorting=none,style=phys-jmr,biblabel=brackets,backref=true,bibencoding=utf8,maxnames=20,eprint=true]{biblatex}
\makeatletter
\pretocmd{\blx@head@bibintoc}{\phantomsection}{}{\ddt}
\makeatother
\DefineBibliographyStrings{english}{%
  backrefpage = {page},
  backrefpages = {pages},
}
\usepackage{csquotes}

\usepackage{pgfplots}
\pgfplotsset{compat=newest}
\usepgfplotslibrary{groupplots}
\usepgfplotslibrary{dateplot}
\newcounter{plot}[figure]
\newcommand{\AddLabel}[1]{%
\node [below right] at (rel axis cs:0,1) {\refstepcounter{plot}\label{#1}\ref{#1})};
}
\usepackage{cleveref}
\crefname{plot}{plot}{plots}
\Crefname{plot}{Plot}{Plots}

\usetikzlibrary{quantikz}
\usepackage{floatrow}
\usepackage[label font=bf,labelformat=simple]{subfig}
\usepackage[font=small,margin=\parindent]{caption}
\usepackage{braket}

\DeclarePairedDelimiter\abs{\lvert}{\rvert}%

\newcommand{\bE}{\mathbf{E}}
\newcommand{\bF}{\mathbf{F}}
\newcommand{\F}{\mathbb{F}}
\newcommand{\bn}{\mathbf{n}}
\newcommand{\bu}{\mathbf{u}}

\newcommand{\by}{\mathbf{y}}

\newcommand{\bc}{\mathbf{c}}
\newcommand{\bs}{\mathbf{s}}

\newcommand{\bH}{\mathbf{H}}

\newcommand{\obu}{\overline{\bu}}

\newcommand{\hbc}{\hat{\mathbf{c}}}
\newcommand{\hbu}{\hat{\mathbf{u}}}
\newcommand{\hbn}{\hat{\mathbf{n}}}
\newcommand{\hu}{\hat{u}}

\newcommand{\oi}{\overline{i}}

\newcommand{\cA}{\mathcal{A}}

\newcommand{\cO}{\mathcal{O}}

\newcommand{\cX}{\mathcal{X}}
\newcommand{\cY}{\mathcal{Y}}

\newcommand{\PM}{\text{PM}}
\newcommand{\PW}{\text{PW}}
\newcommand{\HPW}{\text{HPW}}
\newcommand{\bin}{\text{bin}}
\newcommand{\wt}{\text{wt}}

\renewcommand{\epsilon}{\ensuremath\varepsilon}

\renewcommand{\phi}{\ensuremath{\varphi}}

\usepackage{titling}
\pretitle{\begin{center} \large \bfseries}
\posttitle{\par\end{center}}
\predate{}
\postdate{}
\preauthor{\begin{center} \normalsize}
\postauthor{\end{center}}
\begin{document}

\title{Improved Logical Error Rate via List Decoding of Quantum Polar Codes}

\author{Anqi Gong and Joseph M.~Renes}
\affil{\small Institute for Theoretical Physics, ETH Z\"urich, 8093 Z\"urich, Switzerland}

\date{}
\maketitle

\renewcommand{\abstractname}{\vspace{-1.25\baselineskip}} 
\begin{abstract}
The successive cancellation list decoder (SCL) is an efficient decoder for classical polar codes with low decoding error, approximating the maximum likelihood decoder (MLD) for small list sizes. 
Here we adapt the SCL to the task of decoding quantum polar codes and show that it inherits the high performance and low complexity of the classical case, and can approximate the quantum MLD for certain channels. 
We apply SCL decoding to a novel version of quantum polar codes based on the polarization weight (PW) method, which entirely avoids the need for small amounts of entanglement assistance apparent in previous quantum polar code constructions.
When used to find the precise error pattern, the quantum SCL decoder (SCL-E) shows competitive performance with surface codes of similar size and low-density parity check codes of similar size and rate.  
The SCL decoder may instead be used to approximate the probability of each equivalence class of errors, and then choose the most likely class. 
We benchmark this class-oriented decoder (SCL-C) against the SCL-E decoder and find a noticeable improvement in the logical error rate. 
This improvement stems from the fact that the contributions from just the low-weight errors give a reasonable approximation to the error class probabilities. 
Both SCL-E and SCL-C maintain the complexity $\cO(LN\log N)$ of SCL for code size $N$ and list size $L$.  
We also show that the list decoder can be used to gain insight into the weight distribution of the codes and how this impacts the effect of degenerate errors. 
\end{abstract}
\vspace{1mm}

\section{\label{sec:intro}Introduction}
List decoding, introduced by Elias~\cite{elias_list_1957} and Wozencroft~\cite{wozencroft_list_1958}, is a technique for decoding error-correcting codes in which the decoder is allowed to output a list of possible input codewords instead of a single best guess. 
Also widely studied in the setting of adversarial noise, see e.g.~\cite{guruswami_list_2005,guruswami_essential_2022}, in communication and information theory list decoding is often combined with an existing decoding method and supplemented with a final decoding stage that ensures unique decoding.
For instance, one can simply search the list for the codeword with the smallest distance to the observed channel output. 
The advantage is the list can be used to ameliorate the shortcomings of the original existing decoder without incurring too much computational overhead. 
A notable recent example is the list-enhanced successive cancellation decoder (SCL) proposed by Tal and Vardy~\cite{SCL-CRC}, which considerably boosts the performance of polar codes at the modest multiplicative overhead in complexity linear in the list size.  

In this paper we consider the use of list decoding in the quantum setting. 
More specifically, we study the use of SCL to decode a novel version of quantum polar codes which are based on the recently-developed ``polarization weight'' (PW) and ``higher-order PW'' (HPW) methods for constructing classical polar codes~\cite{PW,HPW}. 
The quantum PW and HPW codes are constructed using the same technique as originally proposed in \cite{QPC}, by combining two classical polar codes. 
When the classical polar codes are constructed according to Ar\i kan's original proposal~\cite{Polar}, that technique nominally produces an entanglement-assisted quantum code, though the amount of assistance can be shown to be small in some cases~\cite{alignment}. 
That suffices for analyzing the capabilities of unassisted codes in the asymptotic limit, but any entanglement-assistance is nonideal for implementing logical qubits. 

Our first contribution is to show that the quantum versions of the PW and HPW construction are CSS codes \cite{CSS-CS,CSS-Steane} requiring zero entanglement assistance. 
We then numerically benchmark the performance of using the classical SCL decoder to decode bit and phase errors separately and find that it gives competitive performance to the surface code of a similar size and to quantum low-density parity-check (QLDPC) codes of similar size and rate, assuming no state preparation and measurement error. 
We call this decoder SCL-E. 
It inherits the computational complexity of the SCL decoder, which is $\cO(LN\log N)$ for code size $N$ and list size $L$. The performance of SCL-E saturates already at a small list size. 
From our experience with numerical simulation, it is the list size used in the classical polar code to approach the maximum likelihood decoding. 
As a rule of thumb, list size $32$ is enough for a PW polar code of blocklength $N=2048$.

Our second contribution is to study the use of the list decoder to determine the error class, not the precise error pattern. 
That is, given the list output, the final decoding stage does not search the list for the lowest-weight error pattern, as in SCL-E, but instead subdivides the list into sets of error patterns which are equivalent modulo the stabilizers of the code, sums the probabilities of the error patterns in each set, and outputs the set with the highest probability.  
We call this decoder SCL-C. Its output determines a logical correction operator, which hopefully matches the logical error on the code incurred by the actual error pattern. 
The complexity overhead relative to SCL-E is only $\cO(LN)$. With a moderate list size (e.g.\ $128$), SCL-C shows a clear improvement in the logical error rate over SCL-E. The improvement continues to grow as the list size gets larger.  Eventually, when the list becomes large enough to contain all the codewords, SCL-C is just the optimal quantum maximum-likelihood decoder (MLD). 
We further use the SCL-C decoder with very large list sizes to explore the weight distribution of low-weight errors in the various error classes, in order to better understand the role of degeneracy. 

The remainder of the paper is organized as follows. 
Section \ref{sec:polar_code} gives a short review of classical polar codes and the successive cancellation list decoder. 
In Section \ref{sec:QEC} we describe our quantum polar code construction method, followed by the details of the SCL-E and SCL-C decoders. 
We present numerical results for the case of independent i.i.d.\ bit and phase noise in Section \ref{sec:simulation}; computer code used to generate these results is publicly available~\cite{gong_gongaa-pw-qpc_2023}. Based on the simulation results, we emphasize the features of SCL decoding that are responsible for the good performance of SCL-E at a small list size and the improved performance of SCL-C in Section \ref{sec:impact_of_list_size}.
We conclude with a discussion of interesting future directions for quantum polar codes, as well as the potential use of list decoding for other codes in Section \ref{sec:conclusion}.

\section{Classical polar code construction and decoding}
\label{sec:polar_code}
\subsection{Construction}
A binary $[N=2^n, K]$ polar code is defined by a polar transform kernel $\bF=\left(\begin{smallmatrix}1 & 0\\1 &1\end{smallmatrix}\right)$ and a set of information indices $\cA=\{i_1,...,i_K\}$; the remaining $N-K$ positions form the ``frozen'' set $\cA^c$. 
The polar encoding circuit is $\bE=\bF^{\otimes n}$, and it encodes an information vector $\bu\in\{0,1\}^N$ into a codeword $\bc=\bu\bE$. The $N-K$ frozen positions of $\bu$ must be fixed to some values (usually all $0$s) that the sender and the receiver have agreed on. 
A useful fact is that $\bE$ is the inverse of itself over $\F_2$, i.e.\ $\bu=\bc\bE$.

A $\emph{polar construction}$ assigns each input a reliability metric and then chooses $K$ of those having the highest metric to form the information set.
In Ar\i kan's original proposal the phenomenon of \emph{channel polarization} is used to determine the information indices. 
Given a binary-input discrete memoryless channel (B-DMC) $W:\cX\triangleq\{0,1\}\to\cY$ with transition probability $W(y|x)$, the polar transform takes $N$ independent instances of $W$ and synthesizes $N$ polarized B-DMCs $W_N^{(i)}$, $0\leq i\leq N-1$.  The channel $W_N^{(i)}:\cX\to\cY^N\times\cX^{i}$ is defined by the transition probability $W_N^{(i)}(\by,\bu_0^{i-1}|u_i)=\tfrac1{2^{N-1}}\sum_{\bu_{i+1}^{N-1}} W_N(\by|\bu)$, where $\bu_i^j$ denotes the sequence $u_i,\dots,u_j$ and $W_N(\by|\bu)=W^{\times N}(\by|\bu\bE)$.
The input indices are chosen to be those whose corresponding synthesized channels have the lowest error probability. 
The synthesized channels will \emph{polarize} in the large $n$ limit, becoming either very reliable or almost completely unreliable. 

More relevant for us are the \emph{polarization weight} (PW)~\cite{PW} and \emph{higher-order polarization weight} (HPW)~\cite{HPW} constructions. To define them, first index the $N=2^n$ rows of $\bE$ from $0$ to $N-1$, starting at the top, and let $\text{bin}(i):=B_{n-1}B_{n-2}...B_1B_0$ be the binary representation of $i$. 
Furthermore define the \emph{$\beta$-expansion} of $i$ for $\beta>0$ as $\bin(i)_{\beta}=\sum_{i=0}^{n-1} B_i\times \beta^i$. 
It is easy to see that $\bin(i)_{2}=i$ and $\bin(i)_{1}=\wt(\bin(i))$, where $\wt(.)$ denotes the Hamming weight.
\begin{definition}\label{def:PW_construction}{\textbf{PW Construction \cite{PW}.}}
Define the polarization weight of row $i$ as 
\setlength\abovedisplayskip{0pt}
\setlength\belowdisplayskip{0pt}
\begin{equation}
\label{eq:PW_def}
    \PW(i)=\bin(i)_{\beta},
\end{equation}
where $\beta$ is chosen to be $2^{1/4}$. 
The $K$ rows with the highest PW are chosen as the information positions.
\end{definition}
\begin{definition}\label{def:HPW_construction}{\textbf{HPW Construction (adapted from \cite{HPW}).} }
Define the higher-order polarization weight of row $i$ as 
\setlength\abovedisplayskip{0pt}
\setlength\belowdisplayskip{0pt}
\begin{equation}
\label{eq:HPW_def}
    \HPW(i)=\sum_{a=1}^{r} c_a\times \bin(i)_{\beta_a},
\end{equation}
for some order $r$ and non-negative constants $c_1=1$, $c_2,...,c_r$ and $\beta_1,...,\beta_r\in [1,2]$. In \cite{HPW}, the parameterization $r=2,c_2=\frac{1}{4},\beta_1=2^{1/4}=\beta_2^4$ is chosen.
\end{definition}
In general, $\beta$ in \eqref{eq:PW_def} should be in the range $[1,2]$ to be consistent with the intuition that $W_N^{(N-1)}$ is the most reliable channel, and $\beta>2$ has the same effect as $\beta=2$. 
In order to break ties, $\PW(i)= \PW(j)$ for $i\neq j$, we can move to a higher order. 
The Reed-Muller code \cite{RM-Muller,RM-Reed} is the $\beta=1$ case in \eqref{eq:PW_def} and an $(r_{RM},n)$-RM code chooses \emph{all} the rows whose index $i$ satisfies $\wt(\bin(i))\geq n-r_{RM}$. 
We make the following modification so that any $K$ rows can be selected.
\begin{definition}\label{def:RM_construction}{\textbf{RM Construction.}}
    The $K$ rows with the highest $\wt(\bin(i))+i/N$ are chosen, i.e.\ $\beta_1=1,\beta_2=2,c_2=1/N$ in \eqref{eq:HPW_def}.
\end{definition}
The PW and HPW polar codes in Definitions \ref{def:PW_construction} and \ref{def:HPW_construction} lead to very good classical codes, and their performance is comparable to the Gaussian approximation (GA) construction on the AWGN channel \cite{PW,beta_expansion,HPW}. 
Our numerical results show that the PW construction family also performs well on the binary symmetric channel (BSC), which is more relevant for the quantum case.

\subsection{Decoding}
Turning to decoding, 
the \emph{successive cancellation decoder} \cite{Polar} is an algorithm to determine $\bu_{\cA}=[u_{i_1},...,u_{i_K}]$ upon receiving a channel output $\by$. 
The decoding is performed successively from $\hu_0$ to $\hu_{N-1}$. 
At step $i$, if a frozen bit is encountered, choose $\hu_i=u_i$, the previously-agreed frozen value. 
Otherwise, the estimation for an information bit is $\hu_i=\text{argmax}_{u_i\in\{0,1\}}W_N^{(i)}(\by,\hbu_0^{i-1}|u_i)$.
This directly estimates the input information vector $\hbu$. The estimate of the corresponding codeword is $\hbc=\hbu\bE$. 

Meanwhile, the SC \emph{list decoder} \cite{SCL-CRC} defines a metric $\PM(\hbu_0^{i})$ for each sequence of decisions $\hbu_0^i$ of $\bu_0^i$, $0\leq i\leq N-1$. 
Each such sequence is called a path, and of course, the decisions for $\bu_{\cA^c}$ must be fixed to the frozen value. The path metric at the end of step $i-1$ is $\PM(\hbu_0^{i-1})\triangleq\Pr[\hbu_0^{i-1}|\by]$. 
After deciding for $u_i$, the path metric is updated to $\PM(\hbu_0^i)=\Pr[\hu_i|\hbu_0^{i-1},\by]\cdot \PM(\hbu_0^{i-1})$.
Note that the path metric $\PM(\hbu_0^{i})$ is related to $W_N^{(i)}(\by,\hbu_0^{i-1}|u_i)$ by a constant, i.e., $W_N^{(i)}(\by,\hbu_0^{i-1}|u_i)=\Pr[\by, \bu_0^{i}]/\Pr[u_i]=2\Pr[\by]\cdot \PM(\hbu_0^{i})$, assuming uniform input.
As in successive cancellation, the list decoder SCL with maximum list size $L$ proceeds from $\hu_0$ to $\hu_{N-1}$. 
At the step of estimating $u_i$, assume there are $L_i$ paths on the list. 
If $u_i$ is a frozen bit, then $\hu_i=u_i$ and all the $L_i$ path metrics are updated according to this decision.
Otherwise $u_i$ is an information bit. 
Instead of picking the $\hu_i$ that maximizes the likelihood $W_N^{(i)}(\by,\hbu_0^{i-1}|u_i)$ as in SC, in SCL each possible choice of $\hat u_i$ is added to each path on the list. 
The list size thus doubles from $L_i$ to $2L_i$. 
If $2L_i>L$, the list is pruned by keeping only the $L$ paths with largest path metric, and SCL proceeds to bit $u_{i+1}$. 
Once all bits are estimated, the decoder outputs the $\hbu$ with the largest path metric, and the corresponding codeword $\hbc=\hbu\bE$.

The SCL decoder has $\cO(LN\log N)$ time complexity and $\cO(LN)$ space complexity. 
The SC decoder is just the SCL decoder with list size $1$.
For implementation details, interested readers can refer to \cite{LLR-SCL} and our source code~\cite{gong_gongaa-pw-qpc_2023}.

The \emph{classical maximum likelihood decoder} (MLD) finds the most likely path, i.e., $\hbu_{\text{ML}}=\text{argmax}_{\bu} W_{N}(\by|\bu)$. 
SCL is a locally greedy algorithm compared to ML because earlier decisions in the path may conflict with a frozen bit encountered subsequently, in that the likelihood of the frozen bit estimated from the path is biased farther away from its actual value. 
Such paths are substantially penalized in their path metric relative to others, but the decoder cannot go back and rule out these suboptimal paths at an earlier stage. 

The SCL decoder described above is a \emph{codeword decoder}, as its output is nominally a codeword. 
For i.i.d.\ BSC, the channel output $\by$ is a noisy codeword $\by=\bu\bE+\bn$ for some noise vector $\bn$. 
The codeword $\hbc=\hbu\bE$ with the largest path metric in the list is the $\hbc$ for which $\wt(\hbn)=\wt(\hbu\bE+\by)$ is the smallest; it will be the output of the SCL decoder.
An alternative approach in this case is \emph{syndrome decoding}. 
Here the goal is to directly find a noise pattern $\hbn$ that is compatible with the observed syndrome $\bs$ and for which $\wt(\hbn)$ is as small as possible.
Notice that $(\bc\bE)_{\cA^c}=\bu_{\cA^c}=\mathbf{0}$ for any codeword $\bc=\bu\bE$ (using the fact that $\bE$ is its own inverse), where $\cA^c$ is the frozen set. 
Therefore, the columns $\cA^c$ of $\bE$ form the parity checks and the syndrome $\bs$ can be obtained via $\bs=(\by\bE)_{\cA^c}=(\bn\bE)_{\cA^c}$. 
To find the most likely noise $\hbn$, put $\bs$ at $\cA^c$ and extend $\bs$ into a size $N$ row vector $\overline{\bs}$, i.e.\ try to decide what values to fill in at the $K$ information bits such that $\overline{\bs}\bE=\hbn$ has as small weight as possible. 
This is essentially the task of giving $00\dots 0$ as a noisy codeword to an SCL codeword decoder that has its frozen values set to $\bs$.

The codeword decoder can of course be used to estimate $\bn$ from $\hbc$ via $\hbn=\hbc+\by$. 
And likewise the syndrome decoder can estimate the input codeword $\bc$ via $\hbc=\hbn+\by$.

\section{Quantum Error Correction}
\label{sec:QEC}
\subsection{Quantum Polar Codes}

Quantum Calderbank-Shor-Steane (CSS) \cite{CSS-CS, CSS-Steane} codes are constructed from two classical codes $C_X$ and $C_Z$ under the requirement $C_Z^\perp\subseteq C_X$ ($C_Z^\perp$ is the dual code to $C_Z$, whose codewords are orthogonal to the codewords of $C_Z$). 
The codewords from $C_X^\perp$ form the $X$-type stabilizers, and the codewords from $C_Z^\perp$ form the $Z$-type stabilizers. Meanwhile, the $X$-type and $Z$-type logical operators are $C_Z\backslash C_X^\perp$ and $C_X\backslash C_Z^\perp$, respectively. 
If $C_X$ is an $[N,K_X,d_X]$ code, and $C_Z$ is an $[N,K_Z,d_Z]$ code, then the resulting quantum code is an $\llbracket N,K_X+K_Z-N, \min(d_X,d_Z)\rrbracket$ code.

A quantum polar code has the same encoding circuit as a classical polar code, except the classical \textsc{cnot} gates  are replaced by quantum \textsc{cnot} gates. 
The resulting $N$-qubit unitary operation is $U=\sum_{\bu\in \{0,1\}^N} \ket{\bu\bE}\bra{\bu}$. 
By design, $U$ implements the encoding circuit $\bE$ in the $Z$ basis. 
By choosing the information set to be $\cA_Z$, we are effectively choosing the rows $\cA_Z$ of $\bE$ to form the generators of $C_Z$. 
The inputs in $\cA_Z^c$ are frozen in the $Z$ basis.  
In the $X$ basis, a simple calculation shows that $U$ acts as $\bE^T=(\bF^T)^{\otimes n}$, i.e.\ $U=\sum_{\obu\in \{\bar{0},\bar{1}\}^N} \ket{\obu\bE^T}\bra{\obu}$, where $\ket{\obu}=\bH^{\otimes N}\ket{\bu}$, $\bH$ is the Hadamard matrix. 
When freezing $\cA^c_X$ in the $X$ basis, rows $\cA_X$ of $\bE^T$ form the generators of $C_X$. 
The code is CSS ($C_Z^\perp\subseteq C_X$) if and only if the $X$ and $Z$ frozen sets are non-intersecting ($\cA^c_X\bigcap \cA^c_Z=\varnothing\Leftrightarrow \cA^c_Z\subseteq \cA_X$).
This is because $C_Z^\perp$ is formed by the columns $\cA^c_Z$ of $\bE$ (for a codeword $\bc=\bu\bE$, it holds that $(\bc\bE)_{\cA^c_Z}=\bu_{\cA^c_Z}=\mathbf{0}$), which is equivalent to being formed by rows $\cA^c_Z$ of $\bE^T$.

The quantum polar code based on the PW construction is a CSS code. A formal definition is the following.

\begin{definition}\label{def:PW_QPC}{\textbf{PW-QPC.}} For $K_X+K_Z> N=2^n$ and $\{i_1,i_2,...,i_N\}$ an ordered set such that $\PW(i_j)>\PW(i_k)$ for $j<k$, the $(N,K_X,K_Z)$ PW quantum polar code is defined by freezing $\{i_{K_Z+1},\dots,i_{N}\}$ in the Z basis and $\{i_1,\dots,i_{N-K_X}\}$ in the X basis.
\end{definition}

In the $Z$ basis, the rows $\{i_{K_Z+1},\dots,i_{N}\}$ are those having the lowest value of PW, hence are frozen. 
In the $X$ basis, the polarization occurs in the reversed direction, i.e.\ $\PW_X(i)=\PW(\oi)$, where $\oi\triangleq N-1-i$, due to the $\bE^T$ action in this basis. 
Using the fact that $\text{bin}(\oi)$ is the bitwise complement of $\text{bin}(i)$ it follows that $\PW(i)+\PW(\oi)=\sum_{j=0}^{n-1}\beta^j$ is constant. 
Therefore $\{i_N,i_{N-1},\dots,i_1\}$ is the ordered set that is decreasing in $\PW_X$, and hence the rows $\{i_1,\dots,i_{N-K_X}\}$ frozen in the $X$ basis correspond precisely to the frozen inputs appropriate for $\PW_X$. 
Since $N-K_X<K_Z+1$, no row is simultaneously frozen in the $X$ and the $Z$ basis. 
The remaining rows $\{i_{N-K_X+1},\dots,i_{K_Z}\}=\cA_X\bigcap\cA_Z$ are used for logical data. 

An HPW-QPC can be defined in the same way, except the ordered set is such that $\HPW(i_j)>\HPW(i_k)$ for $j<k$. The HPW-QPC is also a CSS code since the only fact needed is that $\HPW(i)+\HPW(\oi)$ is constant for all $i$. 
Without further specification, the PW-QPC and HPW-QPC used later are instantiated with the construction parameters in  Definitions \ref{def:PW_construction} and \ref{def:HPW_construction}.

\begin{figure}[hbt]
\includegraphics[width=0.95\textwidth]{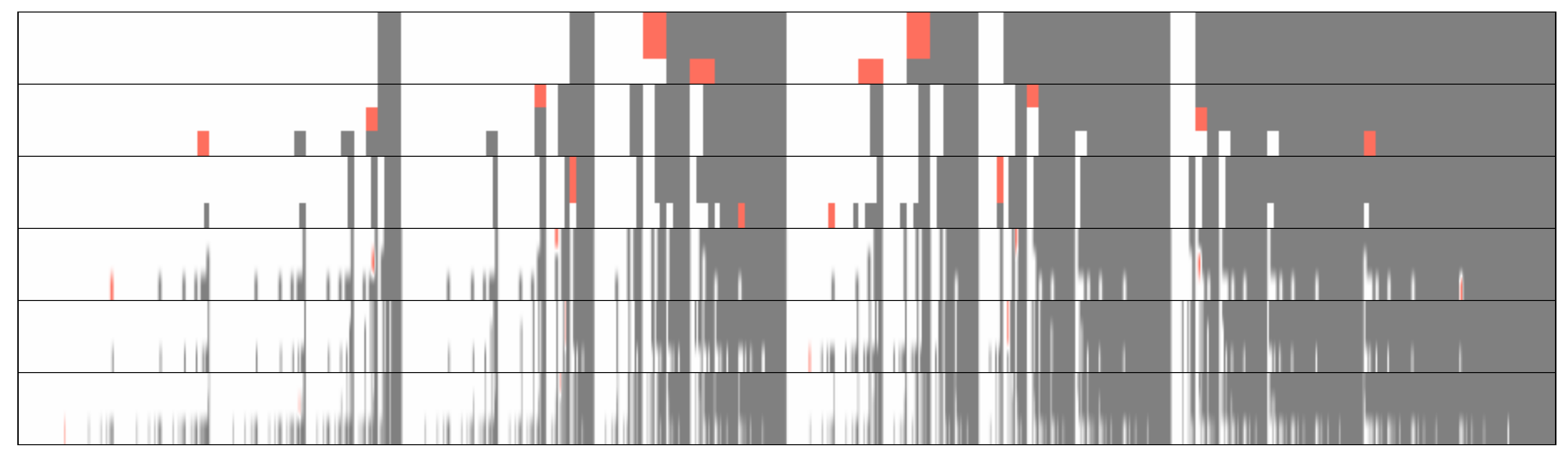}
\caption{\label{fig:info_pos} The position of the two information bits (red) of the families of $K_Z=K_X=N/2+1$ quantum polar code under PW, HPW, RM constructions. 
From top to bottom: $N=64$ to $N=2048$, and left to right positions 0 to $N-1$. In gray are the $X$-basis frozen bits, white the $Z$-basis frozen bits. 
The RM construction has a lot of mixing between the  frozen bits in the two bases, and therefore it is the hardest to decode among the three, when using an SCL decoder with the same list size.}
\end{figure}

An RM-QPC can also be defined using Definition \ref{def:RM_construction}. 
It is one extreme of the polarization weight family ($\beta=1$). 
Table \ref{tab:table_info_pos} records the input indices and minimum distances for various PW, HPW, and RM codes encoding two qubits. 
Figure~\ref{fig:info_pos} shows the positions of frozen bits in the $X$ and $Z$-basis in these codes. 
The two kinds of frozen bits are highly interleaved or mixed together in RM codes, and much less so in PW codes. 
\begin{table}[h]
\caption{\label{tab:table_info_pos}%
Information positions of the two information bits for the $K_Z=K_X=N/2+1$ quantum polar code under the PW, HPW, RM construction, and the distance (the minimum weight of the logical operators) of each code with different sizes. 
Rows are indexed from $0$ to $N-1$.}
\renewcommand{\arraystretch}{1.2}
{\small 
\begin{tabular}{|c|c|c|c|c|}
\hline
\multirow{2}{*}{\textbf{N}} & \multicolumn{3}{c|}{\textbf{Information positions}} & \multirow{2}{*}{\textbf{Distance}}\\
\cline{2-4}
& \textbf{PW} & \textbf{HPW} & \textbf{RM} &\\
\hline
64 & 26,37 & 26,37 & 28,35 & 8\\ \hline
128 & 43,84 & 29,98 & 15,112 & 8\\ \hline
256 & 92,163 & 92,163 & 120,135 & 16\\ \hline
512 & 179,332 & 118,393 & 31,480 & 16\\ \hline
1024 & 364,659 & 364,659 & 496,527 & 32\\ \hline
2048 & 723,1324 & 375,1672 & 63,1984 & 32\\ \hline
\end{tabular}
}
\end{table}
A special case of a QPC constructed from the other extreme ($\beta=2$) was already considered in \cite{Q1}:
\begin{definition}\label{def:Q1_construction}
{\textbf{Q1-QPC \cite{Q1}.}}
    A Q1 quantum polar code only encodes one logical qubit: rows $\{0,...,i-1\}$ are frozen in the $Z$ basis, while rows $\{i+1,..., N-1\}$ are frozen in the $X$ basis.
\end{definition}
Since $N$ is even and only one logical qubit is encoded, the number of $X$ and $Z$ stabilizers have to be different. 
When assuming the same $X$ and $Z$ physical error rate, a performance gap in decoding bit and phase flips is unavoidable. The Q1 construction chooses the input $i$ which has the minimal combined logical error rate. 
In terms of the classical polar code Q1 is constructed from, since there are no information bits appearing before the last frozen bit, the successive cancellation decoding is already the classical ML decoding \cite{how_large}. 
This classical polar code, though decoded perfectly, has a much higher frame error rate than a PW/GA classical polar code decoded using a small list size. 
Nevertheless, the Q1 code has a low quantum logical error rate (see the comparison of PW to Q1 in Figure \ref{fig:Q1_vs_PW}). 
We discuss this interesting behavior further in Appendix~\ref{sec:Q1_vs_PW}.

\subsection{Recycling classical decoders}
Both syndrome and codeword variants of SCL can be employed to decode quantum polar codes, and we call the decoder SCL-E in this context. 
Syndrome decoding is the more standard approach, where measurement of the stabilizer operators yields the syndrome value.  
The CSS property of the codes lends itself to decoding the bit-flip and phase-flip error patterns separately, each of which can be handled by the classical SCL decoder. (One may include correlations between the two kinds of errors in joint decoding, but we do not consider this here.)

For each type of error, the most likely noise $\hbn$ estimated by the syndrome SCL decoder is the correction operation to apply. 
The quantum SCL-E decoder is just the SCL decoder, only the criterion for logical error is different.
The SCL decoding results in a frame error if $\hbn$ is not identical to the actual noise $\bn$ ($\hbn\neq\bn$). 
On the other hand, for bit-flip ($X$-type) noise, the SCL-E decoding results in a logical $X$ error if $\hbn$ and $\bn$ do not differ by an $X$-type stabilizer ($\hbn+\bn\notin C_X^{\perp}$). The same applies for phase-flip ($Z$-type) noise. The quantum code decoding only succeeds if neither a logical $X$ nor a logical $Z$ error occurs.

Nominally, codeword decoding seems incompatible with decoding quantum codes, since no noisy codeword is available at the channel output. 
Steane error correction enables this, however, and indeed delivers both $X$ and $Z$-type noisy codewords~\cite{SteaneEC}. 
The $\hbc$ estimated by the SCL codeword decoder from the noisy codeword $\by$ then specifies a correction operation $\hbn=\hbc+\by$. 
Steane error correction also has the advantage that it is fault-tolerant, however we do not pursue this issue further here. 

\subsection{Degeneracy and list decoding}

It is not necessary to find the precise error pattern for a quantum stabilizer code.
The errors compatible with the syndrome are partitioned into equivalence classes of errors, within which the errors only differ by a stabilizer operator. 
Two bit-flip ($X$-type) errors are degenerate for a CSS code when they differ by an $X$-type stabilizer, and similarly for $Z$-type errors. 
For a code of size $K$, the $2^K$ equivalence classes of $X$-type errors are given by $C_Z/C_X^\perp$, the cosets of $C_X^\perp$ in the code $C_Z$; these correspond to the logical $X$ operators. 
Decoding therefore results in a logical $X$ error when the correction operation $\hbn$ and actual error pattern $\bn$ are such that $\hbn+\bn\notin C_X^\perp$. 

The \emph{quantum ML decoder} finds the most likely error equivalence class based on the observed syndrome.
To formulate the probability explicitly, we focus on the case of independent bit and phase flips, where it suffices to decode $X$ and $Z$ errors independently. 
For independent bit flips occurring with identical probability $p_x$ on each qubit, and $\bc_a$ an $X$-type noise having syndrome $\bs$ (measurement results of $Z$-type stabilizers $C_Z^{\perp}$), all elements in $\bc_a+C_Z$ have the same syndrome, and so do elements in the subset $C_a=\bc_a+C_X^\perp$. 
The probability $\pi(C_a)$ of this $X$-type error coset $C_a$ is
\begin{equation}
\label{eq:qML}
    \pi({C_a})=\sum_{\bn_a\in C_a}\Pr[\bn_a]=(1-p_x)^N\sum_{\bn_a\in C_a}\left(\frac{p_x}{1-p_x}\right)^{\wt(\bn_a)}\,.
\end{equation}
The quantum ML decoder chooses the coset $C_a$ with the largest $\pi(C_a)$, and the correction operation is any element from that coset.  

We propose the following means of using the list decoder to include the effects of degeneracy. 
Instead of finding the lowest-weight error on the syndrome decoder's list, simply emulate \eqref{eq:qML} using only the elements of the list. 
That is, instead of summing over all $\bn_a$ in an error coset, include only those elements which appear on the list. 
We call this version of the SCL decoder SCL-C. 
Details of the algorithm are provided in Appendix~\ref{app:sclc}. 

The smaller the weight of $\bn_a$, the larger the term $\Pr[\bn_a]$ contributes to $\pi(C_a)$. It may be expected that SCL-C will outperform SCL-E if the list contains enough low-weight error patterns such that the probability of each coset can be well-approximated. 

\section{Simulation}
\label{sec:simulation}
In this section we present results of numerical simulation of various $\llbracket N,K\rrbracket$ PW-QPC codes decoded with SCL-E and SCL-C decoders.
The codes are constructed using $\beta=2^{1/4}$, unless otherwise specified. 
Throughout we take $K$ even and  $K_X=K_Z=(N+K)/2$ to obtain codes with symmetric $X$ and $Z$ stabilizers.
We employ the independent bit and phase flip model, and report the logical $X$ error rate under bit-flip ($X$-type) noise alone.
For logical error rate $\lesssim 10^{-3}$, $10^6$ samples are used to obtain each data point, otherwise $10^5$ samples are used.

\begin{figure}[htb]
\begin{tikzpicture}
{\small 
\definecolor{blueviolet}{RGB}{138,43,226}
\definecolor{darkgray176}{RGB}{176,176,176}
\definecolor{lightgray211}{RGB}{211,211,211}
\definecolor{darkorange25512714}{RGB}{255,127,14}
\definecolor{dodgerblue}{RGB}{30,144,255}
\definecolor{gold}{RGB}{255,215,0}
\definecolor{limegreen}{RGB}{50,205,50}
\definecolor{steelblue}{RGB}{31,119,180}
\definecolor{teal}{RGB}{0,128,128}

\begin{axis}[
log basis y={10},
tick align=inside,
tick pos=left,
x grid style={lightgray211},
xlabel={Physical error rate \(\displaystyle p\)},
xmajorgrids,
xmode=log,
xmin=0.009, xmax=0.11,
xtick={0.01,0.1},
xticklabels={
  \(\displaystyle {10^{-2}}\),
  \(\displaystyle {0.1}\),
},
minor x tick num = 9,
scaled x ticks=false,
xtick style={color=black},
y grid style={lightgray211},
ylabel={Logical error rate \(\displaystyle P_L\)},
ymajorgrids,
ymin=7e-5, ymax=1,
ymode=log,
ytick style={color=black},
ytick={0.0001,0.001,0.01,0.1,1,10,100},
yticklabels={
  \(\displaystyle {10^{-4}}\),
  \(\displaystyle {10^{-3}}\),
  \(\displaystyle {0.01}\),
  \(\displaystyle {0.1}\),
  \(\displaystyle {1}\),
},
error bars/y dir=both,
error bars/y explicit,
legend pos=outer north east,
legend cell align=left,
]
\addlegendimage{empty legend}
\addlegendentry{$\mathbf{N}$, $\mathbf{L}$}
\addplot [gold, semithick, solid, mark=*, mark size=0.5, mark options={solid}]
table [x=noise, y=scle, y error plus=yerr, y error minus=yerr] {N64E.dat};
\addlegendentry{$64$, $4$}
\addplot [limegreen, semithick, solid, mark=*, mark size=0.5, mark options={solid}]
table [x=noise, y=scle, y error plus=yerr, y error minus=yerr] {N128E.dat};
\addlegendentry{$128$, $8$}
\addplot [teal, semithick, mark=*, mark size=0.5, mark options={solid}]
table [x=noise, y=scle, y error plus=yerr, y error minus=yerr] {N256E.dat};
\addlegendentry{$256$, $8$}
\addplot [dodgerblue, semithick, mark=*, mark size=0.5, mark options={solid}]
table [x=noise, y=scle, y error plus=yerr, y error minus=yerr] {N512E.dat};
\addlegendentry{$512$, $16$}
\addplot [blue, semithick, mark=*, mark size=0.5, mark options={solid}]
table [x=noise, y=scle, y error plus=yerr, y error minus=yerr] {N1024E.dat};
\addlegendentry{$1024$, $16$}
\addplot [blueviolet, semithick, mark=*, mark size=0.5, mark options={solid}]
table [x=noise, y=scle, y error plus=yerr, y error minus=yerr] {N2048E.dat};
\addlegendentry{$2048$, $32$}

\coordinate (insetPosition1) at (rel axis cs:0.1, 0.55);

\end{axis}

\node at (insetPosition1) {\smash{\rlap{\textcolor{white}{\rule{0.17\textwidth}{0.132\textwidth}}}}};
\begin{axis}[
at={(insetPosition1)},
width=0.26\textwidth,
height=0.22\textwidth,
log basis y={10},
semithick,
tick pos=left,
x grid style={lightgray211},
xmajorgrids,
xmin=0.1, xmax=0.11,
xtick={0.103,0.107},
xtick style={color=black},
xticklabel style={
        /pgf/number format/fixed,
        /pgf/number format/precision=5,
        font=\tiny
},
scaled x ticks=false,
y grid style={lightgray211},
ymajorgrids,
ymin=0.14, ymax=0.4,
yminorgrids,
ytick style={color=black},
yticklabel style={
        /pgf/number format/fixed,
        /pgf/number format/precision=5,
        font=\tiny
},
error bars/y dir=both,
error bars/y explicit,
]

\addplot [gold, solid, mark=*, mark size=0.5, mark options={solid}]
table [x=noise, y=scle, y error plus=yerr, y error minus=yerr] {N64E.dat};

\addplot [limegreen, solid, mark=*, mark size=0.5, mark options={solid}]
table [x=noise, y=scle, y error plus=yerr, y error minus=yerr] {N128E.dat};

\addplot [teal, mark=*, mark size=0.5, mark options={solid}]
table [x=noise, y=scle, y error plus=yerr, y error minus=yerr] {N256E.dat};

\addplot [dodgerblue, mark=*, mark size=0.5, mark options={solid}]
table [x=noise, y=scle, y error plus=yerr, y error minus=yerr] {N512E.dat};

\addplot [blue, mark=*, mark size=0.5, mark options={solid}]
table [x=noise, y=scle, y error plus=yerr, y error minus=yerr] {N1024E.dat};

\addplot [blueviolet, mark=*, mark size=0.5, mark options={solid}]
table [x=noise, y=scle, y error plus=yerr, y error minus=yerr] {N2048E.dat};

\end{axis}
}
\end{tikzpicture}
\caption{\label{fig:PW_SCL_E} Logical $X$ error rate of various $\llbracket N,2\rrbracket$ PW-QPC codes under SCL-E decoding with various list sizes $L$. 
The inset is a magnification of the $p\in[0.1,0.11]$ region. 
The $n=\log_2 N$ and the $n+1$ code are closer together when $n$ is even, which accords with their identical distance (cf. Table~\ref{tab:table_info_pos}).}
\end{figure}
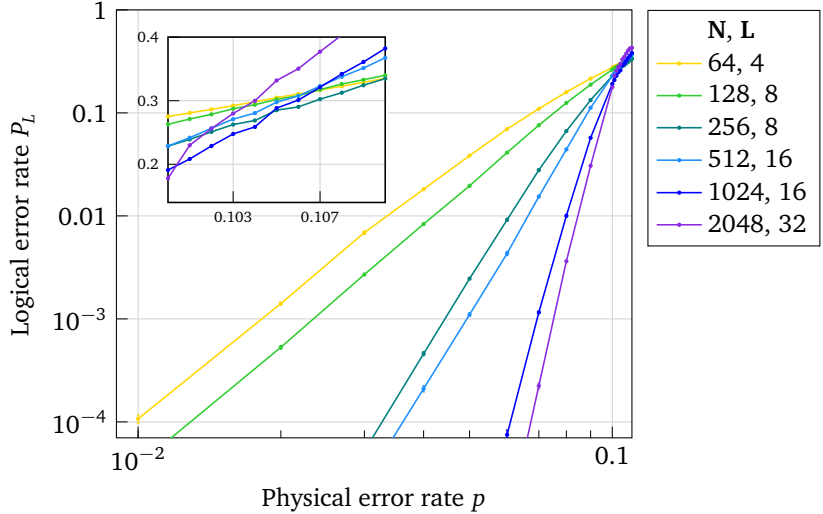

Figure~\ref{fig:PW_SCL_E} shows the SCL-E decoder accuracy for a range of code sizes and noise parameters. 
One can readily see the even/odd $n=\log_2 N$ distinction, just as with the code distance from Table~\ref{tab:table_info_pos}. 
The performance increases considerably when increasing $N$ from $n$ odd to even, but only modestly from even to odd. 

The performance of the $\llbracket 1024,2,32\rrbracket$ PW-QPC under SCL-E decoding (cf. Figure~\ref{fig:PW_SCL_E}) is comparable to the $\llbracket 1201,1,25\rrbracket$ surface code under MLD \cite[Figure 9]{MPS} for $X$-type noise, despite a doubled code rate. 

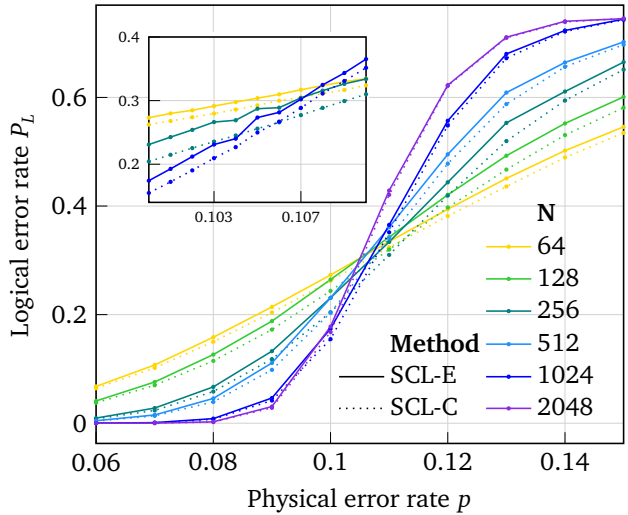
\begin{figure}[htb]
\centering
\begin{tikzpicture}
{\small 
\definecolor{blueviolet}{RGB}{138,43,226}
\definecolor{darkgray176}{RGB}{176,176,176}
\definecolor{lightgray211}{RGB}{211,211,211}
\definecolor{darkorange25512714}{RGB}{255,127,14}
\definecolor{dodgerblue}{RGB}{30,144,255}
\definecolor{gold}{RGB}{255,215,0}
\definecolor{limegreen}{RGB}{50,205,50}
\definecolor{steelblue31119180}{RGB}{31,119,180}
\definecolor{teal}{RGB}{0,128,128}

\begin{axis}[
width=0.5\textwidth,
legend cell align={left},
legend columns=7,
transpose legend,
legend style={
  fill opacity=0,
  draw opacity=1,
  text opacity=1,
  at={(0.97,0.03)},
  anchor=south east,
  draw=none,
  fill=none
},
semithick,
tick pos=left,
x grid style={lightgray211},
xlabel={Physical error rate \(\displaystyle p\)},
xmajorgrids,
xmin=0.06, xmax=0.15,
xtick={0.06,0.08,0.1,0.12,0.14},
xtick style={color=black},
xticklabel style={
        /pgf/number format/fixed,
        /pgf/number format/precision=5
},
scaled x ticks=false,
y grid style={lightgray211},
ylabel={Logical error rate \(\displaystyle P_L\)},
ymajorgrids,
ymin=-0.0372665, ymax=0.77,
ytick style={color=black},
error bars/y dir=both,
error bars/y explicit,
]
\addlegendimage{empty legend}
\addlegendentry{}
\addlegendimage{empty legend}
\addlegendentry{}
\addlegendimage{empty legend}
\addlegendentry{}
\addlegendimage{empty legend}
\addlegendentry{}
\addlegendentry{\textbf{Method}}
\addlegendimage{empty legend}
\addlegendentry{SCL-E}
\addlegendimage{black, line legend}
\addlegendentry{SCL-C}
\addlegendimage{black, line legend, dotted}
\addlegendentry{\textbf{N}}
\addlegendimage{empty legend}

\addplot [gold, mark=*, mark size=0.5, mark options={solid}] table [x=noise, y=scle, y error plus = yerre, y error minus=yerre] {N064.dat};
\addplot [gold, dotted, mark=*, mark size=0.5, mark options={solid}, forget plot] table [x=noise, y=sclc, y error plus = yerrc, y error minus=yerrc] {N064.dat};
\addlegendentry{64}

\addplot [limegreen, mark=*, mark size=0.5, mark options={solid}] table [x=noise, y=scle, y error plus = yerre, y error minus=yerre] {N128.dat};
\addplot [limegreen, dotted, mark=*, mark size=0.5, mark options={solid}, forget plot] table [x=noise, y=sclc, y error plus = yerrc, y error minus=yerrc] {N128.dat};
\addlegendentry{128}

\addplot [teal, mark=*, mark size=0.5, mark options={solid}]
table [x=noise, y=scle, y error plus = yerre, y error minus=yerre] {N256.dat};
\addplot [teal, dotted, mark=*, mark size=0.5, mark options={solid}, forget plot]
table [x=noise, y=sclc, y error plus = yerrc, y error minus=yerrc] {N256.dat};
\addlegendentry{256}

\addplot [dodgerblue, mark=*, mark size=0.5, mark options={solid}]
table [x=noise, y=scle, y error plus = yerre, y error minus=yerre] {N512.dat};
\addplot [dodgerblue, dotted, mark=*, mark size=0.5, mark options={solid}, forget plot]
table [x=noise, y=sclc, y error plus = yerrc, y error minus=yerrc] {N512.dat};
\addlegendentry{512}

\addplot [blue, mark=*, mark size=0.5, mark options={solid}]
table [x=noise, y=scle, y error plus = yerre, y error minus=yerre] {N1024.dat};
\addplot [blue, dotted, mark=*, mark size=0.5, mark options={solid}, forget plot]
table [x=noise, y=sclc, y error plus = yerrc, y error minus=yerrc] {N1024.dat};
\addlegendentry{1024}

\addplot [blueviolet, mark=*, mark size=0.5, mark options={solid}] table [x=noise, y=scle, y error plus = yerre, y error minus=yerre] {N2048.dat};
\addplot [blueviolet, dotted, mark=*, mark size=0.5, mark options={solid}, forget plot]
table [x=noise, y=sclc, y error plus = yerrc, y error minus=yerrc] {N2048.dat};
\addlegendentry{2048}

\coordinate (insetPosition1) at (rel axis cs:0.1, 0.55);
\end{axis}

\node at (insetPosition1) {\smash{\rlap{\textcolor{white}{\rule{0.17\textwidth}{0.132\textwidth}}}}};
\begin{axis}[
at={(insetPosition1)},
width=0.26\textwidth,
height=0.22\textwidth,
log basis y={10},
semithick,
tick pos=left,
x grid style={lightgray211},
xmajorgrids,
xmin=0.1, xmax=0.11,
xtick={0.103,0.107},
xtick style={color=black},
xticklabel style={
        /pgf/number format/fixed,
        /pgf/number format/precision=5,
        font=\tiny
},
scaled x ticks=false,
y grid style={lightgray211},
ymajorgrids,
ymin=0.14, ymax=0.4,
yminorgrids,
ytick style={color=black},
yticklabel style={
        /pgf/number format/fixed,
        /pgf/number format/precision=5,
        font=\tiny
},
error bars/y dir=both,
error bars/y explicit,
]

\addplot [gold, mark=*, mark size=0.5, mark options={solid}, forget plot] table [x=noise, y=scle, y error plus = yerre, y error minus=yerre] {N64thresh.dat};
\addplot [gold, dotted, mark=*, mark size=0.5, mark options={solid}]
table [x=noise, y=sclc, y error plus = yerrc, y error minus=yerrc] {N64thresh.dat};

\addplot [teal, mark=*, mark size=0.5, mark options={solid}, forget plot]
table [x=noise, y=scle, y error plus = yerre, y error minus=yerre] {N256thresh.dat};
\addplot [teal, dotted, mark=*, mark size=0.5, mark options={solid}]
table [x=noise, dotted, y=sclc, y error plus = yerrc, y error minus=yerrc] {N256thresh.dat};

\addplot [blue, mark=*, mark size=0.5, mark options={solid}, forget plot]
table [x=noise, y=scle, y error plus = yerre, y error minus=yerre] {N1024thresh.dat};
\addplot [blue, dotted, mark=*, mark size=0.5, mark options={solid}]
table [x=noise, y=sclc, y error plus = yerrc, y error minus=yerrc] {N1024thresh.dat};

\end{axis}

}
\end{tikzpicture}
\caption{\label{fig:PW} Performance of the SCL-E (solid) and SCL-C (dotted) decoder on the $\llbracket N, 2\rrbracket$ PW-QPC, list size $128$. The inset is a magnification of the region near the presumed threshold $p\approx 0.105$. Since the intersection points gradually drift to the left, we cannot assert that there is a threshold for this sequence of codes and decoders.}
\end{figure}

We demonstrate the logical error rate improvement of SCL-C over SCL-E decoding in several settings. 
Figure~\ref{fig:PW} depicts the results for PW-QPC codes with fixed $K=2$ and increasing $N$, decoded using list size 128.
The HPW construction gives similar results to PW, thus we omit a separate figure. 
Improvement in the RM construction is depicted in Figure \ref{fig:RM}. 
We also confirm a very slight improvement for the Q1 construction.

\subsection{Impact of list size}
\label{sec:impact_of_list_size}
List size 128 was chosen in Figure~\ref{fig:PW} to make the improvement of SCL-C more visible, but in fact the SCL-E decoder already achieves its decoding performance with a smaller list size (cf. Figure~\ref{fig:PW_SCL_E}). This is the list size that the SCL decoder manages to well approximate the MLD in the classical polar code. Furthermore, the smaller the physical error rate, the smaller the needed list size in the SCL and SCL-E decoders.

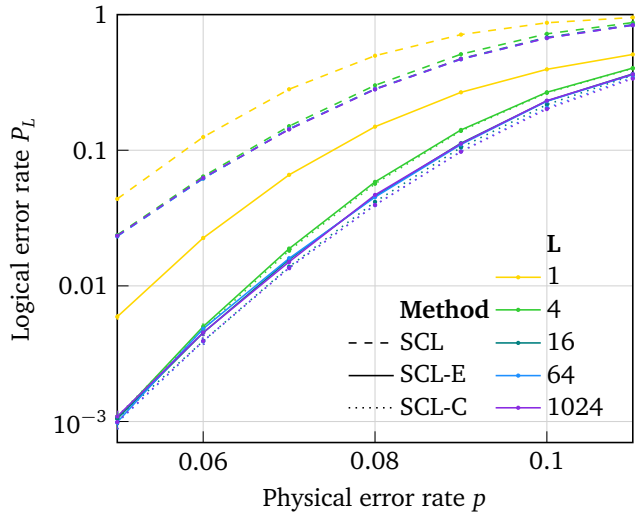
\begin{figure}[htb]
\begin{tikzpicture}
{\small 
\definecolor{lightgray211}{RGB}{211,211,211}
\definecolor{darkgray176}{RGB}{176,176,176}
\definecolor{dodgerblue}{RGB}{30,144,255}
\definecolor{gold}{RGB}{255,215,0}
\definecolor{limegreen}{RGB}{50,205,50}
\definecolor{orchid}{RGB}{218,112,214}
\definecolor{navy}{RGB}{0,0,128}
\definecolor{steelblue31119180}{RGB}{31,119,180}

\definecolor{blueviolet}{RGB}{138,43,226}
\definecolor{darkgray176}{RGB}{176,176,176}
\definecolor{lightgray211}{RGB}{211,211,211}
\definecolor{darkorange25512714}{RGB}{255,127,14}
\definecolor{dodgerblue}{RGB}{30,144,255}
\definecolor{gold}{RGB}{255,215,0}
\definecolor{limegreen}{RGB}{50,205,50}
\definecolor{steelblue31119180}{RGB}{31,119,180}
\definecolor{teal}{RGB}{0,128,128}

\begin{axis}[
legend cell align={left},
legend columns=6,
transpose legend,
legend style={
  fill opacity=0,
  draw opacity=1,
  text opacity=1,
  at={(0.97,0.03)},
  anchor=south east,
  draw=none,
  fill=none
},
semithick,
log basis y={10},
tick align=inside,
tick pos=left,
x grid style={lightgray211},
xlabel={Physical error rate \(\displaystyle p\)},
xmajorgrids,
xmin=0.05, xmax=0.11,
xtick style={color=black},
xtick={0.06,0.08,0.1},
xticklabel style={
        /pgf/number format/fixed,
        /pgf/number format/precision=5
},
scaled x ticks=false,
y grid style={lightgray211},
ylabel={Logical error rate \(\displaystyle P_L\)},
ymajorgrids,
ymin=7e-4, ymax=1,
ytick style={color=black},
ytick={0.001,0.01,0.1,1},
ymode=log,
yticklabels={
  \(\displaystyle {10^{-3}}\),
  \(\displaystyle {0.01}\),
  \(\displaystyle {0.1}\),
  \(\displaystyle {1}\)
},
error bars/y dir=both,
error bars/y explicit,
]
\addlegendimage{empty legend}
\addlegendentry{}
\addlegendimage{empty legend}
\addlegendentry{}
\addlegendentry{\textbf{Method}}
\addlegendimage{empty legend}
\addlegendentry{SCL}
\addlegendimage{black, line legend, dashed}
\addlegendentry{SCL-E}
\addlegendimage{black, line legend}
\addlegendentry{SCL-C}
\addlegendimage{black, line legend, dotted}
\addlegendentry{\textbf{L}}
\addlegendimage{empty legend}

\addplot [gold, dashed, mark=*, mark size=0.5, mark options={solid}, forget plot] table [x=noise, y=scl, y error plus=yerr, y error minus=yerr] {N512L1.dat};
\addplot [gold, mark=*, mark size=0.5, mark options={solid}] table [x=noise, y=scle, y error plus = yerre, y error minus=yerre] {N512L1.dat};
\addlegendentry{1}

\addplot [limegreen, dashed, mark=*, mark size=0.5, mark options={solid}, forget plot] table [x=noise, y=scl, y error plus=yerr, y error minus=yerr] {N512L4.dat};
\addplot [limegreen, mark=*, mark size=0.5, mark options={solid}] table [x=noise, y=scle, y error plus = yerre, y error minus=yerre] {N512L4.dat};
\addplot [limegreen, dotted, mark=*, mark size=0.5, mark options={solid}, forget plot] table [x=noise, y=sclc, y error plus = yerrc, y error minus=yerrc] {N512L4.dat};
\addlegendentry{4}

\addplot [teal, dashed, mark=*, mark size=0.5, mark options={solid}, forget plot] table [x=noise, y=scl, y error plus=yerr, y error minus=yerr] {N512L16.dat};
\addplot [teal, mark=*, mark size=0.5, mark options={solid}] table [x=noise, y=scle, y error plus = yerre, y error minus=yerre] {N512L16.dat};
\addplot [teal, dotted, mark=*, mark size=0.5, mark options={solid}, forget plot] table [x=noise, y=sclc, y error plus = yerrc, y error minus=yerrc] {N512L16.dat};
\addlegendentry{16}

\addplot [dodgerblue, dashed, mark=*, mark size=0.5, mark options={solid}, forget plot] table [x=noise, y=scl, y error plus=yerr, y error minus=yerr] {N512L64.dat};
\addplot [dodgerblue, mark=*, mark size=0.5, mark options={solid}] table [x=noise, y=scle, y error plus = yerre, y error minus=yerre] {N512L64.dat};
\addplot [dodgerblue, dotted, mark=*, mark size=0.5, mark options={solid}, forget plot] table [x=noise, y=sclc, y error plus = yerrc, y error minus=yerrc] {N512L64.dat};
\addlegendentry{64}

\addplot [blueviolet, dashed, mark=*, mark size=0.5, mark options={solid}, forget plot] table [x=noise, y=scl, y error plus=yerr, y error minus=yerr] {N512L1024.dat};
\addplot [blueviolet, mark=*, mark size=0.5, mark options={solid}] table [x=noise, y=scle, y error plus = yerre, y error minus=yerre] {N512L1024.dat};
\addplot [blueviolet, dotted, mark=*, mark size=0.5, mark options={solid}, forget plot] table [x=noise, y=sclc, y error plus = yerrc, y error minus=yerrc] {N512L1024.dat};
\addlegendentry{1024}

\end{axis}

}
\end{tikzpicture}
\caption{\label{fig:PW_list_size} Impact of list size on the $\llbracket 512,2,16\rrbracket$ PW-QPC. Only diminishing gain for SCL-E can be obtained using a list size larger than $16$. While the improvement of SCL-C continues to grow, though tiny, at larger $p$.}
\end{figure}

Figure~\ref{fig:PW_list_size} gives more detail for the case of the $\llbracket 512,2,16\rrbracket$ PW-QPC decoded under varying list sizes. 
Observe that the the performance of classical SCL decoder with $L=16$ is already essentially the same as $L=1024$, and the same holds for the SCL-E decoder. 
The improvement of SCL-C continues to grow with increasing list size, though the gains are small.

It is important to understand why SCL-C makes an improvement and what list size is necessary to see a noticeable effect. 
SCL-C uses all the noise patterns in the list in Equation~\ref{eq:qML} to approximate the probability of each error coset. 
The lowest-weight patterns are the most useful in the approximation, as they constribute significantly to the probability. 
Indeed, the list \emph{heuristically} contains all these noise patterns (if large enough to do so), enabled by the \emph{locally-greedy} nature of the algorithm. 
This explains the improvement of SCL-C at a moderate list size.

Sometimes using \emph{only} the minimum weight patterns may not be a good approximation, especially when $p$ is large. This leads to a performance loss compared to the quantum MLD.
For an equivalence class $E$, let $N_E(w)$ denote the number of error patterns which have Hamming weight $w$. 
To have a good approximation, \cite{MCMC} states that $N_E(w)\cdot \left(\frac{p}{1-p}\right)^w$ should be decreasing with $w$.
However, for the \emph{decision} to be correct, we only need the difference $\abs{N_{E_1}(w)-N_{E_2}(w)}\cdot \left(\frac{p}{1-p}\right)^w$ at the smallest $w$ to be dominating. 

Let $w_{\min}$ and $w_{2,\min}$ denote the smallest and the second smallest $w$ such that $N_E(w)$ is non-zero for some $E$. 
Based on numerical weight distribution results obtained using reasonably large list sizes (at least $1024$) and reported in more detail in Appendix~\ref{sec:examples_of_improvements}, it appears that when $N_{E_1}(w_{\min})\neq N_{E_2}(w_{\min})$, then $\abs{N_{E_1}(w_{\min})-N_{E_2}(w_{\min})}\gg\abs{N_{E_1}(w_{2,\min})-N_{E_2}(w_{2,\min})}\cdot \left(\frac{p}{1-p}\right)^{w_{2,\min}-w_{\min}}$ with high probability, at least up to $N=512$ for $p\lesssim 0.1$. 
This coincides with the logical error rate of the SCL-C curves in Figure \ref{fig:PW_list_size}. 
At $N=512$, list size $64$ is usually enough to fully determine $N_E(w_{\min})$, and list size $1024$ is normally enough to fully determine $N_E(w_{2,\min})$. 
The second-order terms therefore do not matter much, and the improvement is insignificant as we increase the list size from $64$ to $1024$. 
The most perceptable improvement occurs at larger $p$, because there the second-order terms are more relevant.  

Due to the difficulty of analyzing the weight distribution in the error cosets, we do not have any indication whether the assumption that only the low-weight errors matter will hold at larger weights $w$. 
This is also difficult to verify using the list decoder, as the necessary list sizes have very long runtimes. 
However, given the fact that, for the surface code, the optimal decoder only improves the minimum-weight decoder (pair matching) by a factor of $1.4\sim 1.8$ in the two-class decision problem (cf.\ \cite[Figure 9 inset]{MPS}), we suspect that in our $\llbracket N,2\rrbracket$ PW-QPC (four-class decision) the MLD is only able to improve the MWD by a factor of 1.2 in this physical error range. We already achieved a factor of $\sim 1.1$ with SCL-C at only a moderate list size.

\subsection{Rate and error}

\begin{figure}[htb]
\begin{tikzpicture}
{\small
\definecolor{blueviolet}{RGB}{138,43,226}
\definecolor{darkgray176}{RGB}{176,176,176}
\definecolor{lightgray211}{RGB}{211,211,211}
\definecolor{darkorange25512714}{RGB}{255,127,14}
\definecolor{dodgerblue}{RGB}{30,144,255}
\definecolor{gold}{RGB}{255,215,0}
\definecolor{red}{RGB}{255,0,0}
\definecolor{orange}{RGB}{255,93,0}
\definecolor{limegreen}{RGB}{50,205,50}
\definecolor{steelblue31119180}{RGB}{31,119,180}
\definecolor{teal}{RGB}{0,128,128}

\begin{axis}[
width=0.5\textwidth,
log basis y={10},
tick align=inside,
tick pos=left,
x grid style={lightgray211},
xlabel={Physical error rate \(\displaystyle p\)},
xmajorgrids,
xmode=log,
xmin=0.03, xmax=0.11,
xtick={0.03,0.1},
xticklabels={
  \(\displaystyle {0.03}\),
  \(\displaystyle {0.1}\),
},
minor xtick={0.04,0.05,0.06,0.07,0.08,0.09},
xminorgrids=true,
scaled x ticks=false,
xtick style={color=black},
y grid style={lightgray211},
ylabel={Logical error rate \(\displaystyle P_L\)},
ymajorgrids,
ymin=7e-05, ymax=1,
ymode=log,
semithick,
ytick style={color=black},
ytick={0.0001,0.001,0.01,0.1,1,10,100},
yticklabels={
  \(\displaystyle {10^{-4}}\),
  \(\displaystyle {10^{-3}}\),
  \(\displaystyle {0.01}\),
  \(\displaystyle {0.1}\),
  \(\displaystyle {1}\),
},
error bars/y dir=both,
error bars/y explicit,
legend pos=outer north east,
legend cell align=left,
]
\addlegendentry{\textbf{Method}}
\addlegendimage{empty legend}
\addlegendentry{SCL-E}
\addlegendimage{black, line legend}
\addlegendentry{SCL-C}
\addlegendimage{black, line legend, dotted}
\addlegendimage{empty legend}
\addlegendentry{}
\addlegendentry{\textbf{$\boldsymbol{\llbracket}$N,K,d$\boldsymbol{\rrbracket}$, L}}
\addlegendimage{empty legend}

\addplot [gold, mark=*, mark size=0.5, mark options={solid}]
table [x=noise, y=scle, y error plus = yerre, y error minus=yerre] {N64highrate.dat};
\addplot [gold, dotted, mark=*, mark size=0.5, mark options={solid}, forget plot]
table [x=noise, y=sclc, y error plus = yerrc, y error minus=yerrc] {N64highrate.dat};
\addlegendentry{$\llbracket 64,2,8\rrbracket$, $128$}

\addplot [blue, mark=*, mark size=0.5, mark options={solid}]
table [x=noise, y=scle, y error plus = yerre, y error minus=yerre] {N1024K32L128.dat};
\addplot [blue, dotted, mark=*, mark size=0.5, mark options={solid}, forget plot]
table [x=noise, y=sclc, y error plus = yerrc, y error minus=yerrc] {N1024K32L128.dat};
\addlegendentry{$\llbracket 1024,32,16\rrbracket$, $128$}

\addplot [teal, solid, mark=*, mark size=0.5, mark options={solid}]
table [x=noise, y=scle, y error plus=yerr, y error minus=yerr] {N1024K32L16.dat};
\addlegendentry{$\llbracket 1024,32,16\rrbracket$, $16$}

\addplot [limegreen, mark=*, mark size=0.5, mark options={solid}]
table [x=noise, y=scle, y error plus=yerr, y error minus=yerr] {N1024K36L16.dat};
\addlegendentry{$\llbracket 1024,36,16\rrbracket$, $16$}

\addplot [blueviolet, mark=*, mark size=0.5, mark options={solid}]
table [x=noise, y=scle, y error plus=yerr, y error minus=yerr] {N1024K38L16.dat};
\addlegendentry{$\llbracket 1024,38,8\rrbracket$, $16$}

\addplot [orange, mark=*, mark size=0.5, mark options={solid}]
table [x=noise, y=scle, y error plus=yerr, y error minus=yerr] {N1024K42L16beta16921.dat};
\addlegendentry{$\llbracket 1024,42,16\rrbracket$, $16$, $\beta=2^{1/4}-0.02$}

\addplot [red, mark=*, mark size=0.5, mark options={solid}]
table [x=noise, y=scle, y error plus=yerr, y error minus=yerr] {N1024K42L16beta06921.dat};
\addlegendentry{$\llbracket 1024,42,32\rrbracket$, $16$, $\beta=2^{1/4}-0.12$}

\end{axis}
}
\end{tikzpicture}
\caption{\label{fig:PW_high_rate} Logical $X$ error rate of higher-rate PW-QPC codes. 
The $\llbracket 1024,32,16\rrbracket$ and  $\llbracket 64,2,8\rrbracket$ codes have the same rate, $\nicefrac{1}{32}$, but the former has a lower logical error rate than the latter the physical error rate is low. 
A slight improvement of SCL-C over SCL-E still can be seen in the blue curves. 
We reduced the $\beta$ by $0.02$ in the orange curve so as to maintain the distance and to form a fair comparison to the rate $\nicefrac{1}{25}$ random QLDPC in \cite[Figure 5]{roffe_decoding_2020}. Further reducing $\beta$ to $2^{1/4}-0.12$ yields distance 32. 
The resulting red curve performs even better at lower noise; for instance $P_L\approx 4.2\times 10^{-6}$ at $p=0.04$ ($10^7$ samples).}
\end{figure}
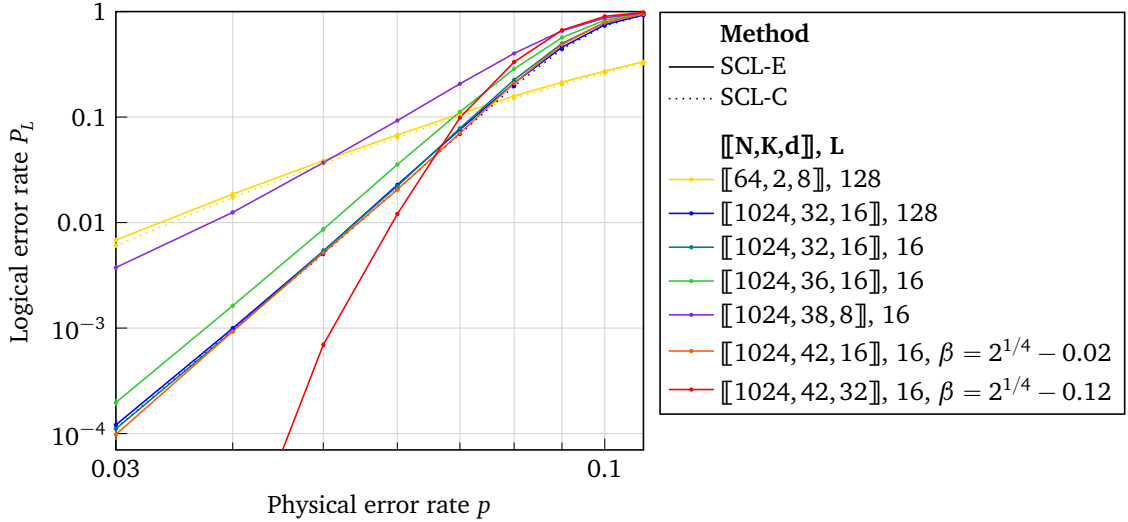

Just as its classical counterpart can be used for high-rate communication, the quantum polar code with a non-vanishing rate has reasonable performance. 
Figure~\ref{fig:PW_high_rate} depicts the logical error rate of the $\llbracket 1024,32,16\rrbracket$ and the $\llbracket 64,2,8\rrbracket$ codes, both with rate $\nicefrac{1}{32}$. 
As expected, a larger blocklength leads to a smaller logical error rate at low physical error rates. 

The codes also compare favorably to constant-rate QLDPC codes constructed by taking the hypergraph product of randomly generated classical LDPC codes and decoded via the BP+OSD-CS method, as reported in \cite[Figure 5]{roffe_decoding_2020}.
However, in trying to emulate the rate $\nicefrac{1}{25}$ of those codes more closely, we observed a significant loss of performance when increasing $K$ from $36$ to $38$ ($N=1024$) because the distance of the code is halved. 
One way to increase the code rate while maintaining the distance is to resort to a slightly smaller $\beta$ for the PW code, at the expense of \emph{potentially} doubling the list size. This is illustrated by the orange curve in our Figure~\ref{fig:PW_high_rate}, where we decrease $\beta$ to $2^{1/4}-0.02$ so that the code has distance $16$. 
In the red curve, where we further decrease $\beta$ to $2^{1/4}-0.12$ so that the code has distance 32, we observe a large performance gain at $p<0.066$, though decoder accuracy is lost in the higher physical error range. 
If extremely high rate is needed at a small physical error rate, the Reed-Muller construction (smallest $\beta$) can be used, as the distance stays at $32$ even when $K$ becomes $252$. 
The decoding complexity of the RM-QPC is not an issue anymore at very small physical error rate, because a small list size suffices there. 
For example, the $\llbracket 1024,252,32\rrbracket$ RM-QPC has logical error rate beneath $10^{-5}$ at $p\leq 0.01$, even when decoded with list size $4$. 
This demonstrates the flexibility of the code family.

\section{Conclusion}
\label{sec:conclusion}
In this work, we introduced the polarization weight construction family of quantum polar code (PW-QPC) that provably satisfies the CSS constraint (no channel is simultaneously frozen in the X and Z basis) and has good logical error rate performance if decoded using a small list size (SCL-E). 
We also show that combining the codewords in the list heuristically (SCL-C) leads to a noticeable performance improvement, because the SCL decoder allows us to recover the lowest-weight noise patterns. 
The downside of this method is that this effect is only noticeable at moderate list size.
From an implementation perspective, the improved performance of SCL-C over SCL-E is probably not worth the increased computational complexity. 
However, from a theoretical point of view, the class-based decoding of SCL-C provides more insight into the role of degeneracy in quantum error-correction. 
We are still lacking a general understanding of how much performance improvements degeneracy can offer, and for what kinds of codes. 
The ability of SCL-C to explore the weight distribution of low-weight errors compatible with the observed syndrome provides a useful tool in exploring this issue. 

There are several avenues for interesting further research.  
Most importantly, the fault tolerance of this family of error correction codes ought to be further investigated, for which \cite{Q1,Goswami_thesis} would be a good starting point. 
Secondly, to optimize the PW-QPC, instead of choosing $\beta=2^{1/4}$ as we did in this paper, $\beta$ can be optimized at each $(N,K,L,p)$ in order to minimize the SCL-E error. As can be seen by comparing the $K=38$ line with the two $K=42$ lines in Figure~\ref{fig:PW_high_rate}, the potential performance gain is quite large.  
Thirdly, the task of tailoring the code and decoder to correlated $X$ and $Z$ noise (e.g., for the depolarizing channel) is left to future work. 

It would also be interesting to investigate whether list decoding can improve the logical error rate for other codes. 
For instance, the implicit list in higher-order BP+OSD type decoding \cite{panteleev_degenerate_2021,roffe_decoding_2020} of quantum low-density parity-check codes might be exploited in the same way as SCL-C.

\section*{Acknowledgments}
We thank Henry D.\ Pfister for useful discussions. JMR acknowledges support from the ETH Quantum Center and the Swiss National Science Foundation Sinergia
grant CRSII5\_186364. 
Numerical simulations were performed on the ETH Zürich Euler cluster.

\appendix
\section{Technical details}
\label{app:sclc}
Upon receiving a noisy codeword $by$, the SCL decoder implementation enables us to access both the list of information decision vectors $\hbu$ and the corresponding codewords $\hbc$ at the same time. 
To partition the codewords in the list into equivalence classes, we only need to look at the $K$ information positions of $\hbu$, namely $\hbu_{\cA_X\bigcap\cA_Z}$. 
Hence we can compute $\wt(\hbc+\by)$ and insert it to the corresponding class according to $\hbu_{\cA_X\bigcap\cA_Z}$ in time $\cO((N+K)L)$. 
Computing the probability of each error coset and making a decision takes time $\cO(L+K)$. 
The total space overhead is $\cO(L)$. Compared to the SCL time and space complexity $\cO(LN\log N)$ and $\cO(LN)$, these overheads are small.


In SCL-E decision-making, if there are two codewords having the same distance to the noise but belonging to different classes, we simply choose randomly. 
However, if the two classes also have the same probability, we ensure that the SCL-C guess is consistent with that of  SCL-E, in order to avoid the fluctuations caused by making uncorrelated random decisions.  

\section{Examples of SCL-C improvements}
\label{sec:examples_of_improvements}
In this section, we give the weight distribution (WD) simulation results for some $\llbracket N,2\rrbracket$ PW-QPC of small blocklength $N$, in order to more fully illustrate why SCL-C can improve SCL-E.
Each error coset can be indexed using the $K$ information bits, i.e., $i\in\{0,...,2^K-1\}$. 
Denote by $N_i(w)$ the number of elements in coset $i$ that have a Hamming weight $w$. 
Let $w^{\min}_i$ be the smallest $w$ so that $N_i(w)\neq 0$, and $w^{\min}$ be the smallest of all $w^{\min}_i$. 
Two cosets $i,j$ have the same WD if $N_i(w)=N_j(w),\ \forall w$. 
The syndrome SCL decoder is very likely to find all the low-weight noise patterns and thus is useful in finding $N_i(w)$ for the small $w$. 

We performed quantum ML decoding on the $(16,9,9)$ PW-QPC and found there is no improvement over the MW decoding. 
This is because this code is highly symmetric. 
Whenever there is was a unique $i$ such that $w^{\min}_i=w^{\min}$, this coset also had the largest probability. 
Whenever there were multiple $i$'s satisfying $w^{\min}_i=w^{\min}$, these cosets all had the same WD and thus the same probability.

Starting from $N=32$, we found the PW-QPC to be less symmetric. 
At $N=32$, we use a list size of $1024$ and assume that cosets are likely to have the same WD (later verified using the full list $2^{17}$) in order to understand the cases where only one of SCL-C and SCL-E makes the correct decision. 
A large portion of them are due to the following situation: Two cosets $i,j$ both have minimal weight elements, i.e.\ $w_i^{\min}=w_j^{\min}=w^{\min}$, and the same weight distribution, but due to limited list size, their observed WD differ at some $w$. SCL-C and SCL-E are basically both random guessing the coset, but these fluctuations are self-canceling on average. 
The remaining cases in which SCL-C decodes correctly and SCL-E incorrectly all have the same pattern: $w^{\min}_i=w^{\min},\forall i$, even cosets have the same WD, odd as well, but $N_{\text{even}}(w_{\min})\neq N_{\text{odd}}(w_{\min})$. 
For example $N_0(5)=N_2(5)=4$, $N_1(5)=N_3(5)=2$, $N_0(7)=N_2(7)=60$, $N_1(7)=N_3(7)=46$. 
Through a probability calculation, one can see that SCL-C has a slightly larger advantage in guessing the correct class.

At $N=64$, more asymmetric cases where SCL-C makes an improvement appear. For example, using list size $1024$, we observe both first-order difference: e.g., $N_1(5)=2, N_3(5)=1, N_1(7)=18, N_3(7)=13, N_0(9)=48, N_2(9)=16$ and second-order difference: e.g., $N_i(9)=8,\forall i$, $N_0(11)=N_2(11)=184$, $N_1(11)=N_3(11)=216$.
With a smaller list size (e.g., $128$), it is usually enough to fully capture the elements contributing to $N_i(w^{\min})$ and simply sample from those contributing to $N_i$ of the second smallest $w$ in proportion to the true values.

\section{Comparison to Q1}
\label{sec:Q1_vs_PW}
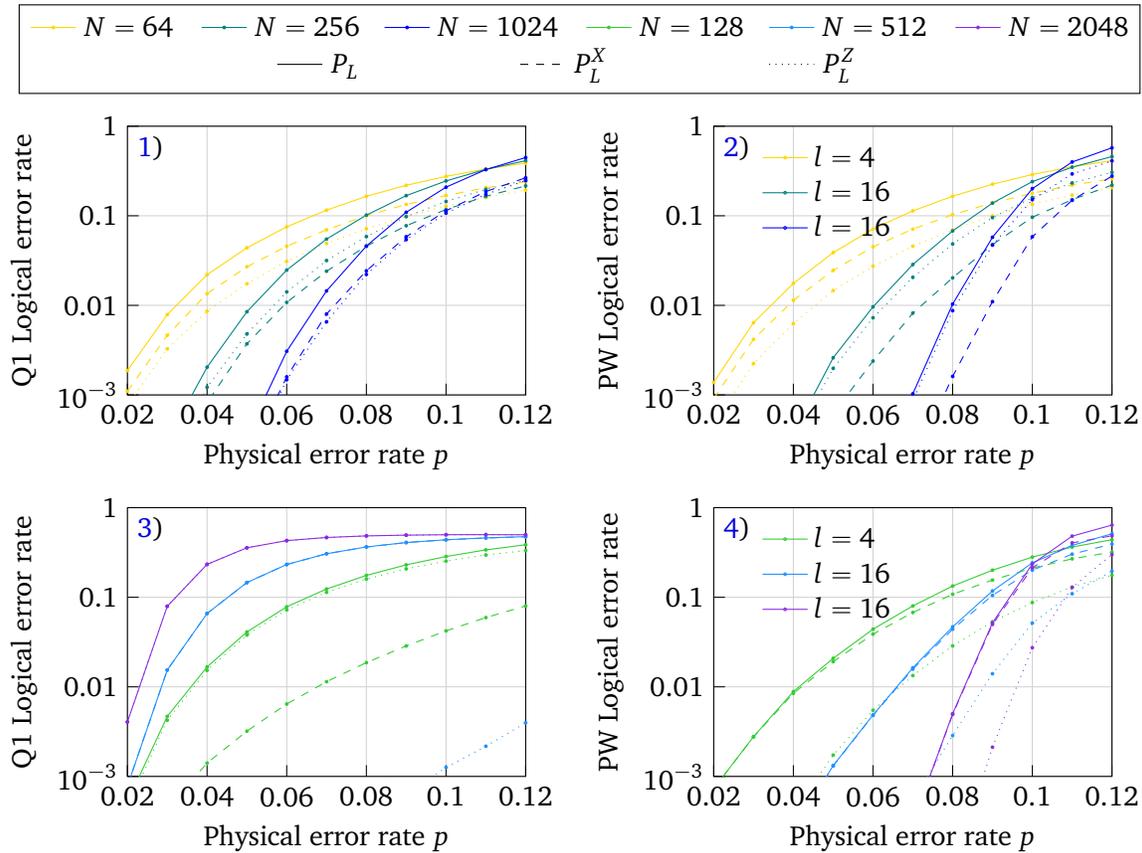
\begin{figure*}[htb]
\begin{tikzpicture}
\definecolor{lightgray211}{RGB}{211,211,211}
\definecolor{gold}{RGB}{255,215,0}
\definecolor{teal}{RGB}{0,128,128}
\definecolor{blueviolet}{RGB}{138,43,226}
\definecolor{dodgerblue}{RGB}{30,144,255}
\definecolor{limegreen}{RGB}{50,205,50}

\begin{groupplot}[
group style={
  group name=myplot,
  group size= 2 by 2, 
  vertical sep=1.5cm,
  horizontal sep=2.5cm
},
width=0.4\textwidth,
height=0.3\textwidth,
legend cell align={left},
legend columns=4,
transpose legend,
legend style={
  fill opacity=0,
  draw opacity=1,
  text opacity=1,
  at={(0.1,0.97)},
  anchor=north west,
  draw=none,
  fill=none
},
log basis y={10},
tick align=inside,
tick pos=left,
x grid style={lightgray211},
xlabel={Physical error rate \(\displaystyle p\)},
xmajorgrids,
xmin=0.02, xmax=0.12,
xtick={0.02,0.04,0.06,0.08,0.1,0.12},
xtick style={color=black},
xticklabel style={
        /pgf/number format/fixed,
        /pgf/number format/precision=5
},
scaled x ticks=false,
xtick style={color=black},
ylabel={Q\(\displaystyle 1\) Logical error rate},
y grid style={lightgray211},
ymajorgrids,
ymin=0.001, ymax=1,
ymode=log,
ytick style={color=black},
ytick={0.001,0.01,0.1,1},
yticklabels={
  \(\displaystyle {10^{-3}}\),
  \(\displaystyle {0.01}\),
  \(\displaystyle {0.1}\),
  \(\displaystyle {1}\)
}
]

\nextgroupplot[ylabel={Q\(\displaystyle 1\) Logical error rate}]
\AddLabel{plot:Q1_even}
\addplot [gold, mark=*, mark size=0.5, mark options={solid}]
table [x=noise, y=pl] {N64Q1.dat};
\label{plots:N=64}
\addplot [gold, dotted, mark=*, mark size=0.5, mark options={solid}, forget plot]
table [x=noise, y=plz] {N64Q1.dat};
\addplot [gold, dashed, mark=*, mark size=0.5, mark options={solid}, forget plot]
table [x=noise, y=plx] {N64Q1.dat};

\addplot [teal, mark=*, mark size=0.5, mark options={solid}]
table [x=noise, y=pl] {N256Q1.dat};
\label{plots:N=256}
\addplot [teal, dotted, mark=*, mark size=0.5, mark options={solid}, forget plot]
table [x=noise, y=plz] {N256Q1.dat};
\addplot [teal, dashed, mark=*, mark size=0.5, mark options={solid}, forget plot]
table [x=noise, y=plx] {N256Q1.dat};

\addplot [blue, mark=*, mark size=0.5, mark options={solid}]
table [x=noise, y=pl] {N1024Q1.dat};
\label{plots:N=1024}
\addplot [blue, dotted, mark=*, mark size=0.5, mark options={solid}, forget plot]
table [x=noise, y=plz] {N1024Q1.dat};
\addplot [blue, dashed, mark=*, mark size=0.5, mark options={solid}, forget plot]
table [x=noise, y=plx] {N1024Q1.dat};

\nextgroupplot[ylabel={PW Logical error rate}]
\AddLabel{plot:PW_even}
\addplot [gold, mark=*, mark size=0.5, mark options={solid}]
table [x=noise, y=pl] {N64K1PW.dat};
\addplot [gold, dotted, mark=*, mark size=0.5, mark options={solid}, forget plot]
table [x=noise, y=plz] {N64K1PW.dat};
\addplot [gold, dashed, mark=*, mark size=0.5, mark options={solid}, forget plot]
table [x=noise, y=plx] {N64K1PW.dat};
\addlegendentry{\(\displaystyle l=4\)}

\addplot [teal, mark=*, mark size=0.5, mark options={solid}]
table [x=noise, y=pl] {N256K1PW.dat};
\addplot [teal, dotted, mark=*, mark size=0.5, mark options={solid}, forget plot]
table [x=noise, y=plz] {N256K1PW.dat};
\addplot [teal, dashed, mark=*, mark size=0.5, mark options={solid}, forget plot]
table [x=noise, y=plx] {N256K1PW.dat};

\addlegendentry{\(\displaystyle l=16\)}

\addplot [blue, mark=*, mark size=0.5, mark options={solid}]
table [x=noise, y=pl] {N1024K1PW.dat};
\addplot [blue, dotted, mark=*, mark size=0.5, mark options={solid}, forget plot]
table [x=noise, y=plz] {N1024K1PW.dat};
\addplot [blue, dashed, mark=*, mark size=0.5, mark options={solid}, forget plot]
table [x=noise, y=plx] {N1024K1PW.dat};
\addlegendentry{\(\displaystyle l=16\)}

\nextgroupplot[ylabel={Q\(\displaystyle 1\) Logical error rate}]
\AddLabel{plot:Q1_odd}
\addlegendimage{black, line legend}
\label{plots:P_L}
\addlegendimage{black, line legend, dashed}
\label{plots:P_L_X}
\addlegendimage{black, line legend, dotted}
\label{plots:P_L_Z}

\addplot [limegreen, mark=*, mark size=0.5, mark options={solid}]
table [x=noise, y=pl] {N128Q1.dat};
\label{plots:N=128}
\addplot [limegreen, dotted, mark=*, mark size=0.5, mark options={solid}, forget plot]
table [x=noise, y=plz] {N128Q1.dat};
\addplot [limegreen, dashed, mark=*, mark size=0.5, mark options={solid}, forget plot]
table [x=noise, y=plx] {N128Q1.dat};

\addplot [dodgerblue, mark=*, mark size=0.5, mark options={solid}]
table [x=noise, y=pl] {N512Q1.dat};
\label{plots:N=512}
\addplot [dodgerblue, dotted, mark=*, mark size=0.5, mark options={solid}, forget plot]
table [x=noise, y=plz] {N512Q1.dat};
\addplot [dodgerblue, dashed, mark=*, mark size=0.5, mark options={solid}, forget plot]
table [x=noise, y=plx] {N512Q1.dat};

\addplot [blueviolet, mark=*, mark size=0.5, mark options={solid}]
table [x=noise, y=pl] {N2048Q1.dat};
\label{plots:N=2048}
\addplot [blueviolet, dotted, mark=*, mark size=0.5, mark options={solid}, forget plot]
table [x=noise, y=plz] {N2048Q1.dat};
\addplot [blueviolet, dashed, mark=*, mark size=0.5, mark options={solid}, forget plot]
table [x=noise, y=plx] {N2048Q1.dat};

\nextgroupplot[ylabel={PW Logical error rate}]
\AddLabel{plot:PW_odd}
\addplot [limegreen, mark=*, mark size=0.5, mark options={solid}]
table [x=noise, y=pl] {N128K1PW.dat};
\addplot [limegreen, dotted, mark=*, mark size=0.5, mark options={solid}, forget plot]
table [x=noise, y=plz] {N128K1PW.dat};
\addplot [limegreen, dashed, mark=*, mark size=0.5, mark options={solid}, forget plot]
table [x=noise, y=plx] {N128K1PW.dat};
\addlegendentry{\(\displaystyle l=4\)}

\addplot [dodgerblue, mark=*, mark size=0.5, mark options={solid}]
table [x=noise, y=pl] {N512K1PW.dat};
\addplot [dodgerblue, dotted, mark=*, mark size=0.5, mark options={solid}, forget plot]
table [x=noise, y=plz] {N512K1PW.dat};
\addplot [dodgerblue, dashed, mark=*, mark size=0.5, mark options={solid}, forget plot]
table [x=noise, y=plx] {N512K1PW.dat};

\addlegendentry{\(\displaystyle l=16\)}
\addplot [blueviolet, mark=*, mark size=0.5, mark options={solid}]
table [x=noise, y=pl] {N2048K1PW.dat};
\addplot [blueviolet, dotted, mark=*, mark size=0.5, mark options={solid}, forget plot]
table [x=noise, y=plz] {N2048K1PW.dat};
\addplot [blueviolet, dashed, mark=*, mark size=0.5, mark options={solid}, forget plot]
table [x=noise, y=plx] {N2048K1PW.dat};
\addlegendentry{\(\displaystyle l=16\)}

\end{groupplot}

\path (myplot c1r1.outer north west)
      (myplot c1r2.outer south west)
;
\path (myplot c1r1.north west|-current bounding box.north)--
      coordinate(legendpos)
      (myplot c2r1.north east|-current bounding box.north);
\matrix[
matrix of nodes,
anchor=south,
draw,
inner sep=0.2em,
draw,
column 1/.style={anchor=base east},
column 2/.style={anchor=base},
column 3/.style={anchor=base west}
]at([yshift=1ex,xshift=-3ex]legendpos)
{
\ref{plots:N=64} $N=64$\quad
\ref{plots:N=256} $N=256$&[5pt]
\ref{plots:N=1024} $N=1024$\quad 
\ref{plots:N=128} $N=128$&[5pt]
\ref{plots:N=512} $N=512$\quad
\ref{plots:N=2048} $N=2048$\\
\ref{plots:P_L} $P_L$&[5pt]
\ref{plots:P_L_X} $P_L^X$&[5pt]
\ref{plots:P_L_Z} $P_L^Z$\\
};
\end{tikzpicture}
\caption{Comparison of the $\llbracket N,1\rrbracket$ Q1-QPC and PW-QPC. Logical X-error rate $P_L^X$ (dashed), logical Z-error rate $P_L^Z$ (dotted), combined logical error rate $P_L=1-(1-P_L^Z)(1-P_L^X)$ (solid). \Cref{plot:Q1_even,plot:PW_even} $n$ even. \Cref{plot:Q1_odd,plot:PW_odd} $n$ odd. List size 1 (SC decoding) is already the MWD for the Q1-QPC. The logical error rates of the PW plots (right column) are the results of the SCL-E decoder and the list sizes used for each blocklength are shown in the figures. The Q1 construction (information position for ignoring correlations) follows from \cite{Q1} Table 1. This position may not be the best for this physical error rate range $p\in[0.02,0.12]$, but the resulting Q1 code is already comparable to PW with a small list size.}

\label{fig:Q1_vs_PW}
\end{figure*}
A comparison between the Q1-QPC and the PW-QPC is shown in Figure \ref{fig:Q1_vs_PW}. 
The information bit for the Q1 construction is taken from \cite[Table 1]{Q1}, and it should be noted that this choice of the information bit may not be optimal in the physical error rate range $[0.02,0.12]$. 
However, the comparison is not intended to show that PW is better than Q1, but instead to highlight the fact that the constituent classical codes in a quantum polar construction need not be good codes themselves. 
Indeed, the performance of Q1 is quite good, and requires no list decoding.  
This leads to the speculation that, maybe a mediocre classical code that can be decoded ML already leads to a decent quantum code. 
On the other hand, how good a classical code at its full potential is, of course, has an influence on the quantum logical error rate lower bound. 
However, due to the computational constraint, a classical good code (like Reed-Muller) may only be decoded far from optimum (no known decoder for $K\geq N/2$ in the Reed-Muller case gives good performance), and the corresponding quantum code may appear to be bad (Figure \ref{fig:RM}).

\emph{Remark: }$K$ being odd is not ideal for the PW-QPC if the error is symmetric. As the blocklength gets larger, the logical error rate $P_L=P_L^X+P_L^Z-P_L^X P_L^Z\approx \max\{P_L^X, P_L^Z\}$. Hence the $\llbracket N, 1\rrbracket$ code only improves the logical error rate of the $\llbracket N, 2\rrbracket$ code by around a factor of $2$. 
While the $\llbracket 2N, 2\rrbracket$ code improves the $\llbracket N, 2\rrbracket$ code by a much larger amount.

\section{Reed-Muller}
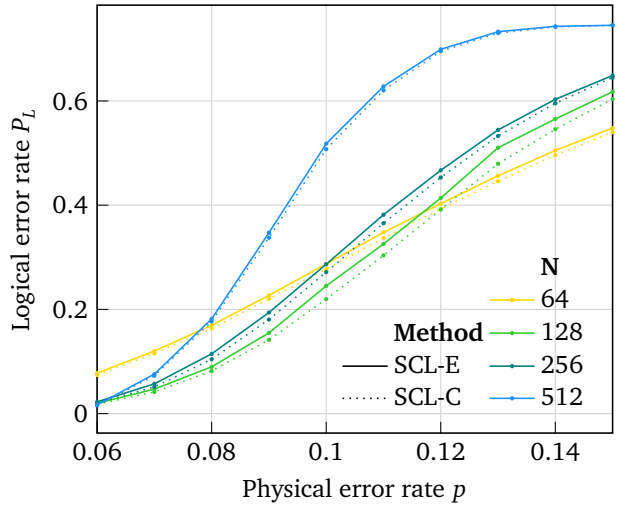
\begin{figure}[hbt]
\begin{tikzpicture}
{\small 
\definecolor{darkgray176}{RGB}{176,176,176}
\definecolor{lightgray211}{RGB}{211,211,211}
\definecolor{dodgerblue}{RGB}{30,144,255}
\definecolor{gold}{RGB}{255,215,0}
\definecolor{limegreen}{RGB}{50,205,50}
\definecolor{teal}{RGB}{0,128,128}

\begin{axis}[
legend cell align={left},
legend columns=5,
transpose legend,
legend style={
  fill opacity=0,
  draw opacity=1,
  text opacity=1,
  at={(0.97,0.03)},
  anchor=south east,
  draw=none,
  fill=none
},
tick align=inside,
tick pos=left,
semithick,
x grid style={lightgray211},
xlabel={Physical error rate \(\displaystyle p\)},
xmajorgrids,
xmin=0.06, xmax=0.15,
xtick={0.06,0.08,0.1,0.12,0.14},
xtick style={color=black},
xticklabel style={
        /pgf/number format/fixed,
        /pgf/number format/precision=5
},
scaled x ticks=false,
xtick style={color=black},
y grid style={lightgray211},
ylabel={Logical error rate \(\displaystyle P_L\)},
ymajorgrids,
ymin=-0.0373274005934481, ymax=0.78387541246241,
ytick style={color=black}
]
\addlegendimage{empty legend}
\addlegendentry{}
\addlegendimage{empty legend}
\addlegendentry{}
\addlegendentry{\textbf{Method}}
\addlegendimage{empty legend}
\addlegendentry{SCL-E}
\addlegendimage{black, line legend}
\addlegendentry{SCL-C}
\addlegendimage{black, line legend, dotted}
\addlegendentry{\textbf{N}}
\addlegendimage{empty legend}
\addplot [gold, mark=*, mark size=0.5, mark options={solid}]
table [x=noise, y=scle, y error plus = yerre, y error minus=yerre] {N64RM.dat};
\addplot [gold, dotted, mark=*, mark size=0.5, mark options={solid}, forget plot]
table [x=noise, y=sclc, y error plus = yerrc, y error minus=yerrc] {N64RM.dat};
\addlegendentry{64}

\addplot [limegreen, mark=*, mark size=0.5, mark options={solid}]
table [x=noise, y=scle, y error plus = yerre, y error minus=yerre] {N128RM.dat};
\addplot [limegreen, dotted, mark=*, mark size=0.5, mark options={solid}, forget plot]
table [x=noise, y=sclc, y error plus = yerrc, y error minus=yerrc] {N128RM.dat};
\addlegendentry{128}

\addplot [teal, mark=*, mark size=0.5, mark options={solid}]
table [x=noise, y=scle, y error plus = yerre, y error minus=yerre] {N256RM.dat};
\addplot [teal, dotted, mark=*, mark size=0.5, mark options={solid}, forget plot]
table [x=noise, y=sclc, y error plus = yerrc, y error minus=yerrc] {N256RM.dat};
\addlegendentry{256}

\addplot [dodgerblue, mark=*, mark size=0.5, mark options={solid}]
table [x=noise, y=scle, y error plus = yerre, y error minus=yerre] {N512RM.dat};
\addplot [dodgerblue, dotted, mark=*, mark size=0.5, mark options={solid}, forget plot]
table [x=noise, y=sclc, y error plus = yerrc, y error minus=yerrc] {N512RM.dat};
\addlegendentry{512}

\end{axis}
}
\end{tikzpicture}
\caption{\label{fig:RM} Logical error rate of the SCL-E (solid) and the SCL-C (dotted) decoder for the $\llbracket N,2\rrbracket$ RM-QPC, list size $128$. $N>512$ is not included as those codes perform very poorly under the SCL decoding on this physical error range.}
\end{figure}

The classical Reed-Muller code is notoriously difficult to decode when the order $r_{RM}$ is large. Due to the large mixing factor (see Figure \ref{fig:info_pos}), using the same list size, the SCL error is higher in RM than in the PW polar code. 
Furthermore, as illustrated in Figure~\ref{fig:RM}, both the improvement of SCL-E over SCL and SCL-C over SCL-E are smaller in the RM-QPC, possibly due to the larger weight of the stabilizers.
Note that the quantum RM code we considered here is different from the quantum Reed-Muller (QRM) codes in the literature, which is a CSS construction from two shortened-RM codes $\overline{\text{RM}}(r,m)$ and $\overline{\text{RM}}(m-r-1,m)$.

\printbibliography[heading=bibintoc,title={\large References}]

\end{document}